\documentclass[12pt]{article}

\usepackage{amsmath,amssymb,amsfonts,amsthm,hyperref,bbm,latexsym,epsfig}
\usepackage{graphicx,multirow}

\newcommand{\bs}{\begin{subequations}}
\newcommand{\es}{\end{subequations}}
\newcommand{\be}{\begin{equation}}
\newcommand{\ee}{\end{equation}}
\newcommand{\ba}{\begin{eqnarray}}
\newcommand{\ea}{\end{eqnarray}}
\newcommand{\no}{\nonumber \\}

\newcommand{\zz}{\mathbbm{Z}}

\newcommand{\diag}{\mbox{diag}}

\allowdisplaybreaks

\usepackage{xcolor}

\begin{document}

\title{
\normalsize \hfill CFTP/16-010
\\[4mm]
\LARGE Group-theoretical search for rows or columns \\
of the lepton mixing matrix}

\author{
Darius~Jur\v{c}iukonis$^{(1)}$\thanks{E-mail: \tt darius.jurciukonis@tfai.vu.lt}
\ and
Lu\'\i s~Lavoura$^{(2)}$\thanks{E-mail: \tt balio@cftp.tecnico.ulisboa.pt}
\\*[3mm]
$^{(1)} \! $
\small Institute of Theoretical Physics and Astronomy, Vilnius University, \\
\small Saul\.etekio ave.\ 3, LT-10222 Vilnius, Lithuania
\\[2mm]
$^{(2)} \! $
\small CFTP, Instituto Superior T\'ecnico, Universidade de Lisboa, \\
\small Av.\ Rovisco Pais 1, 1049-001 Lisboa, Portugal
\\*[2mm]
}

\date{\today}

\maketitle

\begin{abstract}
We have used the {\tt SmallGroups} library of groups,
together with the computer algebra systems {\tt GAP} and {\tt Mathematica},
to search for groups with a three-dimensional irreducible representation
in which one of the group generators has a twice-degenerate eigenvalue
while another generator has non-degenerate eigenvalues.
By assuming one of these group generators
to commute with the charged-lepton mass matrix
and the other one to commute with the neutrino (Dirac) mass matrix,
one derives group-theoretical predictions
for the moduli of the matrix elements of either a row or a column
of the lepton mixing matrix.
Our search has produced several realistic predictions
for either the second row, or the third row,
or for any of the columns of that matrix.
\end{abstract}

\newpage

\section{Introduction}

Over the past few years---starting with ref.~\cite{lam}---a paradigm
has been developed in which the moduli of the entries of the lepton mixing
(Pontecorvo--Maki--Nakagawa--Sakata or PMNS) matrix $U$
are fixed by some discrete,
finite,
non-Abelian group $G$.
In this paradigm,
the (unspecified) family theory of the leptons
is symmetric under $G$.
That group breaks,
through some unspecified mechanism,
into two different subgroups $G_\ell \subset G$ and $G_\nu \subset G$,
which only intersect at the identity element of $G$.
The mass matrices $M_\ell$ and $M_\nu$
of the charged leptons and of the neutrinos,
respectively,
are separately invariant under $G_\ell$ and $G_\nu$,
respectively.
As a consequence,
the unitary matrices that diagonalize $M_\ell$ and $M_\nu$,
named $U_\ell$ and $U_\nu$,
respectively,
also diagonalize the matrices of the restrictions to $G_\ell$ and $G_\nu$,
respectively,
of a three-dimensional representation of $G$.
The diagonalization of $M_\ell$ and of $M_\nu$ is thus replaced
by the diagonalization of the matrices representing
the two subgroups $G_\ell$ and $G_\nu$ in some representation of $G$.
The numerical values of the moduli of the matrix elements
of $U = U_\ell^\dagger U_\nu$ are traced back in this way
to a three-dimensional representation of a finite group $G$.
The irreducible representations of finite groups are finite in number
and may be systematically studied.
That study may be carried out exclusively through
theoretical means~\cite{hagedorn},
but is nowadays greatly facilitated by the free availability
of the computer software {\tt GAP}~\cite{GAP},
which manipulates groups and their representations,
and of the complete library {\tt SmallGroups}
of all the discrete groups of order less than 2\,000~\cite{SG}.

The paradigm mentioned in the previous paragraph
has firstly been developed under the assumption
that the neutrinos are Majorana fields.
In that case,
the group $G_\nu$ must be (isomorphic to)
the Klein group $\zz_2 \times \zz_2$,\footnote{We have assumed
in this statement that $G_\nu$ is a subgroup of $SU(3)$.
If $G_\nu$ is a subgroup of $U(3)$ but not of $SU(3)$,
then it should be of the form $\zz_2 \times \zz_2 \times \zz_2$.}
because the neutrino Majorana masses must remain real and positive
under a rephasing of the neutrino fields
and therefore this rephasing can at most be a change of sign.
Systematic searches using {\tt GAP}
were produced under this assumption~\cite{lam2,lim}
and a thorough classification of the PMNS matrices
achievable in this way has been derived~\cite{fonseca}.
The paradigm has been extended
to the cases of Dirac neutrinos~\cite{hagedorn} and of quarks~\cite{quarks};
then,
$G_\nu$ may be a general $\zz_n$ group with $n > 2$.
An extensive theoretical as well as {\tt SmallGroups} investigation
of those cases has been presented in ref.~\cite{chinese}.

The papers mentioned in the previous paragraph aimed at
fixing the whole matrix $\left| U \right|^2$,
defined as
$\left( \left| U \right|^2 \right)_{ij} \equiv \left| U_{ij} \right|^2$,
through group theory.
In this paper,
following ref.~\cite{lavouraludl},
we have the more modest aim of only fixing either one row
or one column of $\left| U \right|^2$.
This allows the prediction of two out of
the four parameters
of $\left| U \right|^2$.\footnote{Those four parameters
are usually taken to be three mixing (`Euler') angles and one phase,
but they may alternatively be chosen to be
four of the entries of $\left| U \right|^2$~\cite{Branco}.}
This happens when either $G_\ell$ or $G_\nu$,
respectively,
is represented by twice-degenerate matrices,
\textit{i.e.}\ by $3 \times 3$ matrices which have two equal eigenvalues
while the third eigenvalue is different.
For instance,
if the matrices representing $G_\ell$ have two of their eigenvalues equal,
then only one of columns of their diagonalizing matrix $U_\ell$ is well-defined,
\textit{i.e.}\ defined but for an arbitrary overall phase.
Choosing that column to be the third one,
the third row of $U = U_\ell^\dagger U_\nu$ is well-defined
while the first and second rows may be mixed and cannot,
therefore,
be predicted.
In ref.~\cite{lavouraludl} this job of predicting either one row
or one column of $\left| U \right|^2$ was undertaken
under the assumption that the neutrinos are Majorana fields;
it was found that most rows/columns thus found have some zero entries
and are therefore of no practical interest,
since the phenomenology indicates the absence of zeros from the PMNS matrix.

In this paper we shall assume instead that
neutrinos are Dirac fields;\footnote{We thus treat the lepton sector
in exactly the same way as the quark sector,
just as was done in ref.~\cite{chinese}.
We drop any attempt to explain the smallness of the neutrino masses
through the usual see-saw mechanism.
Other versions to that mechanism,
like for instance a see-saw mechanism for the vacuum expectation values
of Higgs doublets~\cite{radovcic}, or a see-saw mechanism
with extra vector-like neutrinos~\cite{valle}, may possibly be employed.}
this allows for a larger variety of groups $G_\ell $ and $G_\nu$---namely,
they will either be cyclic groups of order larger than 2
or possibly groups $\zz_m \times \zz_n$---and consequently
to a much larger variety of predictions
for rows/columns of $\left| U \right|^2$.
In section~\ref{theory} we expose the theory behind our group search.
In section~\ref{searches}
we explain how we have used {\tt GAP} to perform the search.
Section~\ref{results} is devoted to the presentation of the rows/columns
that resulted from the search.
Section~\ref{comparison} compares our presumptive rows/columns
to the actual phenomenological values of $\left| U \right|^2$,
checking which rows/columns are realistic.
In section~\ref{conclusions} we make a short summary
of our work.
%%%%% NEW PERIOD
Appendix~\ref{groups} is devoted to some definitions in group theory
and may be skipped by an uninterested reader.

%%%%% NEW PARAGRAPH
We clarify that this paper reports on a pure computational search
made by using {\tt GAP}.
We have made neither any attempt at studying analytically
a particular group or set of groups,
nor at explaining analytically
the results found through our computational search.
For comparison,
ref.~\cite{talbert} gives
other papers relying strongly on the power of the {\tt GAP} software.

\section{Theory} \label{theory}

We work in the context of the three-generation Standard Model
with the addition of three right-handed neutrinos.
The neutrinos are assumed to be standard Dirac fields;
they have no Majorana mass terms.
The charged-lepton mass matrix $M_\ell$
and the neutrino mass matrix $M_\nu$ are defined through the mass terms
\be
\mathcal{L}_\mathrm{lepton\ mass} =
- \overline{\ell_L} M_\ell \ell_R - \overline{\nu_L} M_\nu \nu_R
+ \mathrm{H.c.}
\ee
The matrices $H_\ell \equiv M_\ell M_\ell^\dagger$
and $H_\nu \equiv M_\nu M_\nu^\dagger$ are diagonalized as
\bs
\ba
U_\ell^\dagger H_\ell U_\ell &=&
\diag \left( m_e^2,\ m_\mu^2,\ m_\tau^2 \right),
\\
U_\nu^\dagger H_\nu U_\nu &=&
\diag \left( m_1^2,\ m_2^2,\ m_3^2 \right),
\ea
\es
where $U_\ell$ and $U_\nu$ are $3 \times 3$ unitary matrices.
The PMNS matrix is $U = U_\ell^\dagger U_\nu$.

We assume that the matrices $H_\ell$ and $H_\nu$ are invariant
under the action of two invertible matrices $T_\ell$ and $T_\nu$,
respectively.
This invariance is defined through
\bs
\ba
T_\ell H_\ell T_\ell^{-1} &=& H_\ell, \label{biuytp}
\\
T_\nu H_\nu T_\nu^{-1} &=& H_\nu. \label{buiyo}
\ea
\es
Equation~(\ref{biuytp}) states that $T_\ell$ and $H_\ell$ commute,
therefore they are simultaneously diagonalizable.
Similarly,
equation~(\ref{buiyo}) implies that
$T_\nu$ and $H_\nu$ are simultaneously diagonalizable.
Therefore,
$U_\ell$ diagonalizes $T_\ell$ and $U_\nu$ diagonalizes $T_\nu$:
\bs
\label{buisp}
\ba
U_\ell^\dagger T_\ell U_\ell &=& \hat T_\ell
\ \, \equiv\ \,  \mathrm{diag} \left( l_1,\ l_2,\ l_3 \right), \label{bxihp}
\\
U_\nu^\dagger T_\nu U_\nu &=& \hat T_\nu
\ \, \equiv \ \, \mathrm{diag} \left( n_1,\ n_2,\ n_3 \right).
\ea
\es

We now make the crucial assumption that
$T_\ell$ and $T_\nu$ together generate a
(three-dimensional representation of a)
group which is \emph{finite}.
This assumption allows one to restrict the PMNS matrix,
which in general belongs to the continuous set of unitary matrices,
to a discrete set of matrices defined by the theory of finite groups,
thereby generating some predictive power.
This predictive power is further enhanced if we assume that
the group generated by $T_\ell$ and $T_\nu$ is \emph{small},
\textit{i.e.}\ that its order is smaller than some arbitrary number.
In our practical search
we have assumed that the order of $G$ is smaller than 2\,000,
since this is the present reach of the {\tt SmallGroups} library.

\subsection{Main search}

In our main search we have assumed that two,
and only two,
of the three eigenvalues $l_{1,2,3}$ of $T_\ell$ are equal,
%%%%% $l_1 = l_2 \neq l_3$,
%%%%% while the eigenvalues $n_{1,2,3}$ of $T_\nu$ are all distinct.
while the eigenvalues $n_{1,2,3}$ of $T_\nu$ are all distinct.
Suppose for instance that $l_1 = l_2 \neq l_3$.
Then,
the third column of $U_\ell$,
which according to equation~\eqref{bxihp}
is the normalized eigenvector of $T_\ell$ corresponding to the eigenvalue $l_3$,
is well-defined---only its overall phase is arbitrary---but the first
two columns of $U_\ell$ are not,
because they are eigenvectors corresponding to the same eigenvalue $l_1 = l_2$
and may therefore be arbitrarily mixed between themselves.
As a consequence,
the third row of $U = U_\ell^\dagger U_\nu$ will be fixed
except for its phase,
while the first two rows will remain arbitrary
(they will only be restricted to being orthogonal to the third row
and to each other).
Our assumption thus allows us to `predict' the moduli of the matrix elements
of the third row of $U$.
%%%%% I HAVE ADDED THE FOLLOWING PERIOD.
In the same way,
if $\ell_1 = \ell_3 \neq \ell_2$,
then the second row of $U$ is predicted;
if $\ell_2 = \ell_3 \neq \ell_1$,
then the first row of $U$ is predicted.

In practice,
we compute those moduli in the following way.
Let $p$ and $q$ be two integers,
then,
from equations~\eqref{buisp},\footnote{The use of traces
like those in equation~\eqref{uiyhp}
has been first advocated in ref.~\cite{Branco}.}
\ba
\mathrm{tr} \left( T_\ell^p T_\nu^q \right) &=&
\mathrm{tr} \left( U_\ell \hat T_\ell^p U_\ell^\dagger
U_\nu \hat T_\nu^q U_\nu^\dagger \right)
=
\mathrm{tr} \left( U^\dagger \hat T_\ell^p U \hat T_\nu^q \right)
=
\sum_{i=1}^3 \sum_{j=1}^3  l_i^p n_j^q \left| U_{ij} \right|^2
\no &=&
l_1^p n_1^q \left( 1 - \left| U_{21} \right|^2 - \left| U_{31} \right|^2 \right)
+ l_1^p n_2^q \left( 1 - \left| U_{22} \right|^2 - \left| U_{32} \right|^2 \right)
\no & &
+ l_1^p n_3^q \left( \left| U_{21} \right|^2 + \left| U_{31} \right|^2 
+ \left| U_{22} \right|^2 + \left| U_{32} \right|^2 - 1 \right)
+ l_2^p n_1^q \left| U_{21} \right|^2
\no & &
+ l_2^p n_2^q \left| U_{22} \right|^2
+ l_2^p n_3^q \left( 1 - \left| U_{21} \right|^2 - \left| U_{22} \right|^2 \right)
+ l_3^p n_1^q \left| U_{31} \right|^2
\no & &
+ l_3^p n_2^q \left| U_{32} \right|^2
+ l_3^p n_3^q \left( 1 - \left| U_{31} \right|^2 - \left| U_{32} \right|^2 \right)
\no &=&
\left( l_2^p - l_1^p \right) \left( n_1^q - n_3^q \right)
\left| U_{21} \right|^2 
+ \left( l_2^p - l_1^p \right) \left( n_2^q - n_3^q \right)
\left| U_{22} \right|^2 
\no & &
+ \left( l_3^p - l_1^p \right) \left( n_1^q - n_3^q \right)
\left| U_{31} \right|^2 
+ \left( l_3^p - l_1^p \right) \left( n_2^q - n_3^q \right)
\left| U_{32} \right|^2
\no & &
+ l_1^p n_1^q
+ l_1^p n_2^q
- l_1^p n_3^q
+ l_2^p n_3^q
+ l_3^p n_3^q.
\label{uiyhp}
\ea
Next,
using our extra assumption that $l_1 = l_2 \neq l_3$,
\ba
\mathrm{tr} \left( T_\ell^p T_\nu^q \right) &=&
\left( l_3^p - l_1^p \right) \left( n_1^q - n_3^q \right)
\left| U_{31} \right|^2 
+ \left( l_3^p - l_1^p \right) \left( n_2^q - n_3^q \right)
\left| U_{32} \right|^2
\no & &
+ l_1^p n_1^q
+ l_1^p n_2^q
+ l_3^p n_3^q.
\label{vyurp}
\ea
Writing both $\mathrm{tr} \left( T_\ell T_\nu \right)$
and $\mathrm{tr} \left( T_\ell T_\nu^2 \right)$ as in equation~\eqref{vyurp}
(\textit{i.e.}\ with $p= 1$ and $q = 1$ and $2$, respectively),
one obtains two equations for $\left| U_{31} \right|^2$
and $\left| U_{32} \right|^2$.
They are solved to yield
\bs
\label{sdihp}
\ba
\left| U_{31} \right|^2 &=& \frac{\mathrm{tr} \left( T_\ell T_\nu^2 \right)
+ l_1 \chi - \left( n_2 + n_3 \right) \mathrm{tr} \left( T_\ell T_\nu \right)
- l_1 n_1^2 + l_3 n_2 n_3}
{\left( l_3 - l_1 \right) \left( n_1 - n_2 \right)
\left( n_1 - n_3 \right)},
\label{uigdp} \\
\left| U_{32} \right|^2 &=& \frac{\mathrm{tr} \left( T_\ell T_\nu^2 \right)
+ l_1 \chi - \left( n_1 + n_3 \right) \mathrm{tr} \left( T_\ell T_\nu \right)
- l_1 n_2^2 + l_3 n_1 n_3}
{\left( l_3 - l_1 \right) \left( n_2 - n_1 \right)
\left( n_2 - n_3 \right)},
\label{bihpi} \\
\left| U_{33} \right|^2 &=& \frac{\mathrm{tr} \left( T_\ell T_\nu^2 \right)
+ l_1 \chi - \left( n_1 + n_2 \right) \mathrm{tr} \left( T_\ell T_\nu \right)
- l_1 n_3^2 + l_3 n_1 n_2}
{\left( l_3 - l_1 \right) \left( n_3 - n_1 \right)
\left( n_3 - n_2 \right)},
\label{buigf}
\ea
\es
where $\chi \equiv n_1 n_2 + n_1 n_3 + n_2 n_3$.
(We have used $\left| U_{33} \right|^2 = 1 - \left| U_{31} \right|^2
- \left| U_{32} \right|^2$ to derive equation~\eqref{buigf}
from equations~\eqref{uigdp} and~\eqref{bihpi}.)
Equations~\eqref{sdihp} allow us to compute
the third row of $\left| U \right|^2$
from the matrices $T_\ell$ and $T_\nu$ and from their eigenvalues,
without having to explicitly diagonalize those matrices.

%%%%% NEW TEXT STARTS HERE.

We emphasize that our use of traces $\mbox{tr} \left( T_\ell^p T_\nu^q \right)$
constitutes an important technical progress
over other searches using {\tt GAP},
because {\tt GAP} frequently gives the matrices
of a group representation in non-unitary form;
by using the traces one may directly use those matrices in that form,
without having firstly to unitarize them
and then to compute their eigenvectors;
one thus saves a considerable amount of computer time.

\subsection{Secondary search}

We have also considered the possibility that
there are two matrices $T_{\nu 1}$ and $T_{\nu 2}$
which commute both with each other and with the Hermitian matrix $H_\nu$;
therefore,
they are all diagonalized by the same unitary matrix $U_\nu$:
\bs
\ba
U_\nu^\dagger T_{\nu 1} U_\nu &=& \diag \left( n_1,\ n_2,\ n_3 \right),
\\
U_\nu^\dagger T_{\nu 2} U_\nu &=&
\diag \left( \bar n_1,\ \bar n_2,\ \bar n_3 \right).
\ea
\es
We have considered the situation in which
both $T_{\nu 1}$ and $T_{\nu 2}$ have two identical eigenvalues,
but together they act as if all their eigenvalues are different:
$n_1 = n_2 \neq n_3$ and $\bar n_1 = \bar n_3 \neq \bar n_2$.
We identify this property by computing the invariant quantity
\ba
q &\equiv& 3 \left[ \mathrm{tr} \left( T_{\nu 1} T_{\nu 2} \right) \right]^2
- 2 \left( \mathrm{tr}\, T_{\nu 1} \right) \left( \mathrm{tr}\, T_{\nu 2} \right)
\mathrm{tr} \left( T_{\nu 1} T_{\nu 2} \right)
\no & &
+ \left( \mathrm{tr}\, T_{\nu 1} \right)^2 \mathrm{tr} \left( T_{\nu 2}^2 \right)
+ \mathrm{tr} \left( T_{\nu 1}^2 \right) \left( \mathrm{tr}\, T_{\nu 2} \right)^2
- 3\, \mathrm{tr} \left( T_{\nu 1}^2 \right) \mathrm{tr} \left( T_{\nu 2}^2 \right).
\label{huigp}
\ea
It is easy to check that
$q = - \left( n_1 - n_3 \right)^2
\left( \bar n_1 - \bar n_2 \right)^2 \neq 0$
if $n_1 = n_2$ and $\bar n_1 = \bar n_3$,
while
$q = 0$ if $n_1 = n_2$ and $\bar n_1 = \bar n_2$.
So,
we select $q \neq 0$.

Now,
with
\bs
\ba
U_\ell^\dagger T_\ell U_\ell &=& \diag \left( l_1,\ l_1,\ l_3 \right),
\\
U_\nu^\dagger T_{\nu 1} U_\nu &=& \diag \left( n_1,\ n_1,\ n_3 \right),
\\
U_\nu^\dagger T_{\nu 2} U_\nu &=& \diag \left( \bar n_1,\ \bar n_2,\ \bar n_1
\right),
\ea
\es
and $U = U_\ell^\dagger U_\nu$ as before,
one easily finds that
\bs
\label{huoiy}
\ba
\left| U_{33} \right|^2 &=& \frac{\mathrm{tr} \left( T_\ell T_{\nu 1} \right)
- l_1 n_1 - l_1 n_3 - l_3 n_1}{\left( l_1 - l_3 \right)
\left( n_1 - n_3 \right)},
\\
\left| U_{32} \right|^2 &=& \frac{\mathrm{tr} \left( T_\ell T_{\nu 2} \right)
- l_1 \bar n_1 - l_1 \bar n_2 - l_3 \bar n_1}{\left( l_1 - l_3 \right)
\left( \bar n_1 - \bar n_2 \right)}.
\ea
\es
So,
in this case one can once again derive the entries
in the third row of $\left| U \right|^2$ by using invariant traces.

\section{The search} \label{searches}

We have scanned all the finite groups with order less than $2\,000$
by making use of the library {\tt SmallGroups}.
We have identified which of them
have three-dimensional faithful irreducible representations (`irreps') and,
moreover,
are \emph{not}\/ the direct product of a group $\zz_n$ with $n \ge 2$
by some other group.\footnote{The cyclic group $\zz_n$
is formed by the $n$-roots of unity
under the standard multiplication of complex numbers.
It is of course an Abelian group,
because the multiplication of complex numbers is commutative.}
We only need to scan for \emph{faithful}\/ representations;
if a representation is unfaithful
(\textit{i.e.}\ if it represents several elements of the group $G$
by the same matrix),
then it is the faithful representation of a subgroup $G^\prime \subset G$
and we will find it when we scan for the faithful representations of $G^\prime$.
Moreover,
we only need to scan for \emph{irreducible}\/ representations:
if a representation is reducible,
then all its matrices may simultaneously be rotated to a basis
in which they are all block-diagonal;
hence,
$T_\ell$ and $T_\nu$ will be block-diagonal in some basis,
and therefore $U_\ell$, $U_\nu$,
and the PMNS matrix will also be block-diagonal;
but this is in contradiction with the phenomenology,
which indicates that $U$ has no zero matrix elements.
Finally,
we shed groups $G$ of the form $G = G^\prime \times \zz_n$
with $n \ge 2$,
because a faithful representation of $G^\prime \times \zz_n$
is necessarily also a faithful representation of the smaller group $G^\prime$
and we will find it when we study $G^\prime$.

Since the number of groups of order 1\,536 is much too large
for all of them to be scanned within a reasonable time,
we have used the conjecture in ref.~\cite{chinese}
%%%%% I ADDED A FOOTNOTE
%%%%% that both nilpotent groups
that both nilpotent groups\footnote{One can check
whether the group with {\tt SmallGroups} identifier $\left[ m, n \right]$
($m$ and $n$ are integers;
$m$ is the order of the group)
is nilpotent by first typing the {\tt GAP} command {\tt G:=SmallGroup([m,n])}
and then the {\tt GAP} command {\tt IsNilpotentGroup(G)}.
The latter command produces the answer {\tt True}
if the group $\left[ m, n \right]$ is nilpotent.}
%%%%% I ADDED A FOOTNOTE
%%%%% and groups with a normal Sylow 3-subgroup
and groups with a normal Sylow 3-subgroup\footnote{The Sylow 3-subgroups
of a group $G$ may be found by typing the {\tt GAP} command
{\tt SylowSubgroup(G,3)}.
The {\tt GAP} command {\tt IsNormal(G,U)} returns {\tt True}
if $U$ is a normal subgroup of $G$.}
%%%%% I ADDED A FOOTNOTE
%%%%% never have three-dimensional faithful irreps.
never have three-dimensional faithful irreps.\footnote{In appendix~\ref{groups}
we attempt to explain in simple terms
what are nilpotent groups and groups
with a normal Sylow 3-subgroup~\cite{grouptheory}.}
Since 99.97\% of the groups of order 1\,536 are in one of those two categories,
this conjecture has allowed us
to outright disconsider most of the groups of that order.

Tables~\ref{table1},
\ref{table2},
and~\ref{table3} present all the groups $G$ that we have found
\begin{table}[ht!]
\begin{tabular}{|l|}
\hline
\begin{minipage}[t]{0.96\columnwidth}
{\bf [12, 3]}, [21, 1], {\bf [24, 12]}, [27, 3], [27, 4], {\bf [36, 3]}, 
[39, 1], {\bf [48, 3]}, {\bf [48, 30]}, {\bf [54, 8]}, [57, 1], 
{\bf [60, 5]}, [63, 1], {\bf [75, 2]}, [81, 6], {\bf [81, 7]},
{\bf [81, 8]}, {\bf [81, 9]}, {\bf [81, 10]}, [81, 14],
{\bf [84, 11]}, [93, 1], {\bf [96, 64]}, {\bf [96, 65]}, {\bf [108, 3]}, 
{\bf [108, 11]}, {\bf [108, 15]}, {\bf [108, 19]}, {\bf [108, 21]},
{\bf [108, 22]}, [111, 1], [117, 1], [129, 1], {\bf [144, 3]},
[147, 1], {\bf [147, 5]}, {\bf [150, 5]}, {\bf [156, 14]}, {\bf [162, 10]},
{\bf [162, 12]}, {\bf [162, 14]}, {\bf [162, 44]}, {\bf [168, 42]},
[171, 1], [183, 1], [189, 1],
[189, 4], [189, 5], [189, 7], [189, 8], {\bf [192, 3]}, {\bf [192, 182]},
{\bf [192, 186]}, [201, 1], {\bf [216, 17]}, {\bf [216, 25]},
{\bf [216, 88]}, {\bf [216, 95]}, [219, 1], {\bf [225, 3]}, {\bf [228, 11]}, 
[237, 1], {\bf [243, 16]}, {\bf [243, 19]}, {\bf [243, 20]}, [243, 24], 
{\bf [243, 25]}, {\bf [243, 26]}, {\bf [243, 27]}, [243, 50], 
{\bf [243, 55]}, {\bf [252, 11]}, [273, 3], [273, 4], [279, 1], [291, 1], 
{\bf [294, 7]}, {\bf [300, 13]}, {\bf [300, 43]}, [309, 1], {\bf [324, 3]}, 
{\bf [324, 13]}, {\bf [324, 15]}, {\bf [324, 17]}, {\bf [324, 43]}, 
{\bf [324, 45]}, {\bf [324, 49]}, 
{\bf [324, 50]}, {\bf [324, 51]}, {\bf [324, 60]},
{\bf [324, 102]}, {\bf [324, 111]}, {\bf [324, 128]}, 
[327, 1], [333, 1], {\bf [336, 57]}, [351, 1], [351, 4], [351, 5], [351, 7], 
[351, 8], {\bf [363, 2]}, {\bf [372, 11]}, [381, 1], {\bf [384, 568]}, 
{\bf [384, 571]}, {\bf [384, 581]}, 
[387, 1], [399, 3], [399, 4], [417, 1],
{\bf [432, 3]}, {\bf [432, 33]}, {\bf [432, 57]}, 
{\bf [432, 100]},{\bf [432, 102]}, {\bf [432, 103]},
{\bf [432, 239]}, {\bf [432, 260]}, {\bf [432, 273]},
[441, 1], {\bf [441, 7]}, {\bf [444, 14]}, [453, 1],
{\bf [468, 14]}, [471, 1], {\bf [486, 26]}, 
{\bf [486, 28]}, {\bf [486, 61]}, 
{\bf [486, 125]}, {\bf [486, 164]}, [489, 1], [507, 1], {\bf [507, 5]}, 
[513, 1], [513, 5], [513, 6], [513, 8], [513, 9], {\bf [516, 11]}, 
{\bf [525, 5]}, [543, 1], [549, 1], [567, 1], [567, 4], [567, 5], [567, 7], 
{\bf [567, 12]}, {\bf [567, 13]}, 
{\bf [567, 14]}, {\bf [567, 23]}, [567, 36], {\bf [576, 3]},
[579, 1], {\bf [588, 11]}, {\bf [588, 16]}, {\bf [588, 60]},
[597, 1], {\bf [600, 45]}, {\bf [600, 179]}, [603, 1],
{\bf [624, 60]}, [633, 1], {\bf [648, 19]}, {\bf [648, 21]},
{\bf [648, 23]}, {\bf [648, 244]}, {\bf [648, 259]}, {\bf [648, 260]},
{\bf [648, 266]}, {\bf [648, 352]}, {\bf [648, 531]}, {\bf [648, 532]},
{\bf [648, 533]}, {\bf [648, 551]}, {\bf [648, 563]}, [651, 3], [651, 4],
[657, 1], [669, 1], {\bf [675, 5]}, {\bf [675, 9]}, 
{\bf [675, 11]}, {\bf [675, 12]}, {\bf [684, 11]}, [687, 1], [711, 1],
[723, 1], {\bf [726, 5]}, 
{\bf [729, 62]},{\bf [729, 63]}, {\bf [729, 64]}, {\bf [729, 80]}, 
{\bf [729, 86]}, [729, 94], 
{\bf [729, 95]}, {\bf [729, 96]}, {\bf [729, 97]}, {\bf [729, 98]},
{\bf [729, 284]}, [729, 393], 
{\bf [729, 397]}, {\bf [732, 14]}, [741, 3], [741, 4], {\bf [756, 11]}, 
{\bf [756, 113]}, {\bf [756, 114]}, {\bf [756, 116]}, {\bf [756, 117]}
\end{minipage}\tabularnewline
\hline
\end{tabular}
%%%%% \caption{The {\tt GAP} identifiers of the groups $G$
\caption{The {\tt SmallGroups} identifiers of the groups $G$
with three-dimensional irreducible representations.
Part 1: groups with $\mbox{order}(G) < 768$.
The identifiers in boldface denote the groups which have three-dimensional
irreducible representations in which some of the matrices
have twice-degenerate eigenvalues;
only those groups are relevant for this paper.}
\label{table1}
\end{table}
\begin{table}[ht!]
\begin{tabular}{|l|}
\hline
\begin{minipage}[t]{0.96\columnwidth}
{\bf [768, 1083477]}, {\bf [768, 1085333]}, {\bf [768, 1085335]}, 
{\bf [768, 1085351]}, 
[777, 3], [777, 4], {\bf [804, 11]}, [813, 1], [819, 3], [819, 4],
[831, 1], [837, 1], 
[837, 4], [837, 5], [837, 7], [837, 8], [849, 1], {\bf [864, 69]},
{\bf [864, 194]}, {\bf [864, 675]}, {\bf [864, 701]}, {\bf [864, 703]},
{\bf [864, 737]}, {\bf [867, 2]}, [873, 1],  {\bf [876, 14]}, {\bf [900, 66]},
[903, 5], [903, 6], {\bf [912, 57]}, [921, 1], [927, 1], [939, 1],
{\bf [948, 11]}, {\bf [972, 3]}, {\bf [972, 29]}, {\bf [972, 31]},
{\bf [972, 64]}, {\bf [972, 117]}, {\bf [972, 121]}, {\bf [972, 122]},
{\bf [972, 123]}, {\bf [972, 147]}, {\bf [972, 152]}, {\bf [972, 153]},
{\bf [972, 170]}, {\bf [972, 309]}, {\bf [972, 348]}, {\bf [972, 411]},
{\bf [972, 520]}, {\bf [972, 550]}, {\bf [975, 5]}, [981, 1], [993, 1],
[999, 1], [999, 5], [999, 6], [999, 8], [999, 9], {\bf [1008, 57]},
[1011, 1], {\bf [1014, 7]}, [1029, 6], {\bf [1029, 9]}, [1047, 1], 
[1053, 16], [1053, 25], [1053, 26],  [1053, 2], {\bf [1053, 29]},
{\bf [1053, 32]}, {\bf [1053, 35]}, {\bf [1053, 37]}, [1053, 47],
{\bf [1080, 260]}, [1083, 1], {\bf [1083, 5]}, {\bf [1089, 3]},
{\bf [1092, 68]}, {\bf [1092, 69]}, [1101, 1], {\bf [1116, 11]}, [1119, 1], 
[1137, 1], [1143, 1], [1161, 6], [1161, 9], 
[1161, 10], [1161, 11], [1161, 12], {\bf [1164, 14]}, {\bf [1176, 57]}, 
{\bf [1176, 243]}, [1191, 1], [1197, 3], [1197, 4], 
{\bf [1200, 183]}, {\bf [1200, 384]}, {\bf [1200, 682]}, [1209, 3],
[1209, 4], [1227, 1], {\bf [1236, 11]}, [1251, 1], [1263, 1], [1281, 3],
[1281, 4], {\bf [1296, 3]}, {\bf [1296, 35]}, {\bf [1296, 37]},
{\bf [1296, 39]}, {\bf [1296, 220]}, {\bf [1296, 222]}, {\bf [1296, 226]}, 
{\bf [1296, 227]}, {\bf [1296, 228]}, {\bf [1296, 237]}, {\bf [1296, 647]},
{\bf [1296, 688]}, {\bf [1296, 689]}, {\bf [1296, 699]}, {\bf [1296, 1239]},
{\bf [1296, 1499]}, {\bf [1296, 1995]}, {\bf [1296, 2113]},
{\bf [1296, 2203]}, [1299, 1], {\bf [1308, 14]}, [1317, 1], [1323, 1],
[1323, 4], [1323, 5], [1323, 7], [1323, 8], {\bf [1323, 14]}, {\bf [1323, 40]},
{\bf [1323, 42]}, {\bf [1323, 43]}, {\bf [1332, 14]}, {\bf [1344, 393]},
{\bf [1350, 46]}, [1359, 1], [1371, 1], [1389, 1], {\bf [1404, 14]},
{\bf [1404, 137]},  {\bf [1404, 138]}, {\bf [1404, 140]}, {\bf [1404, 141]},
[1407, 3], [1407, 4], [1413, 1], {\bf [1425, 5]}, [1443, 3], [1443, 4],
{\bf [1452, 11]}, {\bf [1452, 34]}, {\bf [1458, 615]}, {\bf [1458, 618]}, 
{\bf [1458, 659]}, {\bf [1458, 663]}, {\bf [1458, 666]}, {\bf [1458, 1095]}, 
{\bf [1458, 1354]}, {\bf [1458, 1371]}, [1461, 1], [1467, 1], 
{\bf [1488, 57]}, [1497, 1], [1521, 1], {\bf [1521, 7]}, {\bf [1524, 11]}, 
[1533, 3], [1533, 4]
\end{minipage}\tabularnewline
\hline
\end{tabular}
%%%%% \caption{The {\tt GAP} identifiers of the groups
\caption{The {\tt SmallGroups} identifiers of the groups
with three-dimensional irreducible representations.
Part 2: groups $G$ with $768 \le \mbox{order}(G) < 1536$.
The identifiers in boldface stand for groups with three-dimensional
irreducible representations in which some of the matrices
have twice-degenerate eigenvalues.}
\label{table2}
\end{table}
\begin{table}[ht!]
\begin{tabular}{|l|}
\hline
\begin{minipage}[t]{0.96\columnwidth}
{\bf [1536, 408544632]}, {\bf [1536, 408544641]}, {\bf [1536, 408544678]}, 
{\bf [1536, 408544687]}, [1539, 16], [1539, 25], [1539, 26], [1539, 27], 
{\bf [1539, 29]}, {\bf [1539, 32]}, {\bf [1539, 35]}, {\bf [1539, 37]},
[1539, 47], {\bf [1548, 11]}, [1569, 1], {\bf [1575, 7]}, {\bf [1587, 2]},
{\bf [1596, 55]}, {\bf [1596, 56]}, [1623, 1], [1629, 1], [1641, 1],
[1647, 6], [1647, 9], [1647, 10], [1647, 11], [1647, 12], [1659, 3],
[1659, 4], {\bf [1668, 11]}, [1677, 3], [1677, 4], [1701, 68], 
{\bf [1701, 102]}, {\bf [1701, 112]}, {\bf [1701, 115]}, [1701, 126],
[1701, 127], [1701, 128], {\bf [1701, 130]}, {\bf [1701, 131]},
{\bf [1701, 135]}, {\bf [1701, 138]}, [1701, 240], {\bf [1701, 261]},
[1713, 1], {\bf [1728, 3]}, {\bf [1728, 185]}, {\bf [1728, 953]},
{\bf [1728, 1286]}, {\bf [1728, 1290]}, {\bf [1728, 1291]},
{\bf [1728, 2785]}, {\bf [1728, 2847]}, {\bf [1728, 2855]},
{\bf [1728, 2929]}, [1731, 1], {\bf [1734, 5]}, [1737, 1],
{\bf [1764, 11]}, {\bf [1764, 91]}, [1767, 3], [1767, 4],
{\bf [1776, 60]}, [1791, 1], [1803, 1], [1809, 6], [1809, 9], [1809, 10],
[1809, 11], [1809, 12], {\bf [1812, 11]}, [1821, 1], 
[1839, 1], [1857, 1], {\bf [1872, 60]}, {\bf [1875, 16]},
{\bf [1884, 14]}, [1893, 1], [1899, 1], [1911, 3], [1911, 4],
{\bf [1911, 14]}, [1929, 1], {\bf [1944, 35]}, 
{\bf [1944, 37]}, {\bf [1944, 70]}, {\bf [1944, 707]},
{\bf [1944, 746]}, {\bf [1944, 832]}, {\bf [1944, 833]}, {\bf [1944, 849]},
{\bf [1944, 1123]}, {\bf [1944, 2293]}, {\bf [1944, 2294]},
{\bf [1944, 2333]}, {\bf [1944, 2363]}, {\bf [1944, 2415]},
{\bf [1944, 3448]}, [1953, 3], [1953, 4], {\bf [1956, 11]},
[1971, 6], [1971, 9], [1971, 10], [1971, 11], [1971, 12], [1983, 1]
\end{minipage}\tabularnewline
\hline
\end{tabular}
%%%%% \caption{The {\tt GAP} identifiers of the groups
\caption{The {\tt SmallGroups} identifiers of the groups
with three-dimensional irreducible representations.
Part 3: groups with $1536 \le \mbox{order}(G) < 2000$.
The identifiers in boldface denote groups with three-dimensional
irreducible representations with matrices
having twice-degenerate eigenvalues.}
\label{table3}
\end{table}
not to be of type $G = \zz_n \times G^\prime$
and to possess at least one faithful three-dimensional irrep.
Tables~\ref{table1} and~\ref{table2}
reproduce equation~(42) of ref.~\cite{kingludl},
while table~\ref{table3} is new.

We have explicitly constructed all the three-dimensional irreps
of the groups in tables~\ref{table1},
\ref{table2},
and~\ref{table3}.
From each of those irreps
we have discarded matrices proportional to the unit matrix.
We have divided the remaining matrices (of each irrep of each group)
into two sets:
the ones that have non-degenerate eigenvalues (set~$N$)
and the ones that have twice-degenerate eigenvalues (set~$T$).
We recall that a matrix $M$ has degenerate eigenvalues when
\ba
4 \left[ I_2 \left( M \right) \right]^3
- \left( \mathrm{tr}\, M \right)^2 \left[ I_2 \left( M \right) \right]^2
+ 27 \left( \det{M} \right)^2
& & \no
+ 4 \left( \mathrm{tr}\, M \right)^3 \det{M}
- 18 \left( \mathrm{tr}\, M \right) \left[ I_2 \left( M \right) \right]
\det{M} &=& 0,
\ea
where $I_2 \left( M \right) \equiv M_{11} M_{22} + M_{11} M_{33} + M_{22} M_{33}
- M_{12} M_{21} - M_{13} M_{31} - M_{23} M_{32}$ is the second-order invariant of $M$.
We have found that there are many groups for which set $T$ is empty;
we have discarded those groups.
The remaining groups---the ones that
have at least one three-dimensional faithful irrep
with a non-empty set $T$---are marked boldface in tables~\ref{table1},
\ref{table2},
and~\ref{table3}.
Those were the sole relevant groups and irreps for the remainder of our search.

We have then explicitly computed the eigenvalues of all the matrices
in both sets $N$ and $T$.

In our main search,
we have considered all possible pairs of
one matrix $T_\ell$ from set $T$ and one matrix $T_\nu$ from set $N$.
For each of those pairs we have computed the three $\left| U_{3j} \right|^2
(j = 1, 2, 3)$ by using equations~\eqref{sdihp};
in those equations,
$l_1$ is the degenerate eigenvalue of $T_\ell$,
$l_3$ is the non-degenerate eigenvalue of $T_\ell$,
and $n_{1,2,3}$ are the three eigenvalues of $T_\nu$.
We have discarded the set of the three $\left| U_{3j} \right|^2$
whenever any one of them turned out to vanish;
we have only collected the sets for which all three numbers
$\left| U_{31} \right|^2$,
$\left| U_{32} \right|^2$,
and $\left| U_{33} \right|^2$ were non-zero.

For our secondary search we have picked all possible couples
of two matrices $T_{\nu 1}$ and $T_{\nu 2}$ from the set $T$
and selected those couples that commute and that moreover
have a non-zero quantity $q$ defined in equation~\eqref{huigp}.
We have then picked one third matrix $T_\ell$,
and have computed the three $\left| U_{3j} \right|^2$
by using equations~\eqref{huoiy} together with
$\left| U_{31} \right|^2 = 1 - \left| U_{32} \right|^2 - \left| U_{33} \right|^2$.
Once again,
we have discarded all the sets of three $\left| U_{3j} \right|^2$
in which any one of them happened to vanish.

Since all the finite-dimensional representations of finite groups are unitary,
all the representations that we have dealt with are in principle equivalent
to representations through unitary matrices.
{\tt GAP} usually gives the representations in non-unitary form,
but we never have had to bring the representations to unitary form,
because all our computations were performed
in terms of basis-invariant quantities.

At the end of our search we have used {\tt GAP}
to find out the form of the group generated by the matrices $T_\ell$ and $T_\nu$
(in the main search) alone.
Let $\left\langle T_\ell,\ T_\nu \right\rangle$ denote that group.
It coincides in most cases with the initial group $G$,
but sometimes it is just a subgroup of it.

\section{Results} \label{results}

\begin{figure}[h!]
%\centerline{\epsfig{file=figurerows1.eps,width=0.40\textwidth}}
\begin{center}
\includegraphics[scale=0.47]{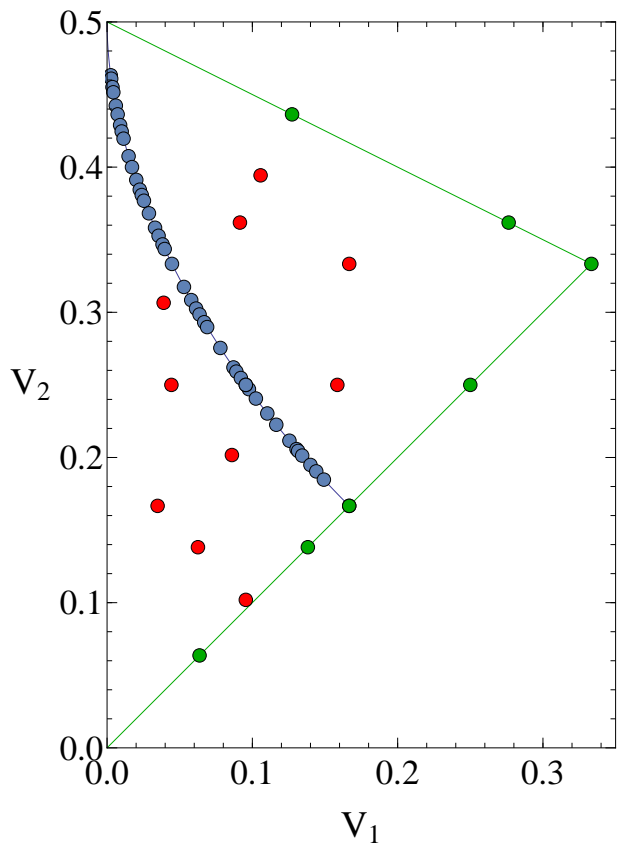}
\end{center}
\vspace*{-3mm}
\caption{A depiction of the sixty structures
$\left\{ V_1,\ V_2,\ V_3 \right\}$ that have been produced by our searches.
The horizontal line gives $V_1$ and the vertical line gives $V_2$,
while $V_3 = 1 - V_1 - V_2$. 
We have discarded structures
in which any of the three $V_j\ (j = 1, 2, 3)$ is zero.
The structures are ordered as $V_1 \le V_2 \le V_3$.
All the points are within the triangle bounded by the vertical axis
(equivalent to $V_1 = 0$) and by the green line.
The latter is composed of two segments with equations
$V_1 = V_2$ and $V_1 + 2 V_2 = 1\, \Leftrightarrow\, V_2 = V_3$.
The two segments meet at the point $V_1 = V_2 = V_3 = 1/3$.
The blue curve unites points with similar features (see text).}
\vspace*{3mm}
\label{figura}
\end{figure}

The searches described in the previous section
have produced a total of sixty sets
of three non-zero numbers $\left| U_{31} \right|^2$,
$\left| U_{32} \right|^2$,
and $\left| U_{33} \right|^2$;\footnote{Actually,
all the sets except one have been produced by the main search.
All but one of the sets produced by the secondary search merely reproduce sets
that had already been obtained in the main search.}
in each set,
$\left| U_{31} \right|^2 + \left| U_{32} \right|^2
+ \left| U_{33} \right|^2 = 1$.
From now on,
we let $V_j \equiv \left| U_{3j} \right|^2\ (j = 1, 2, 3)$
denote the three numbers in each set
and we assume that they have been ordered as $V_1 \le V_2 \le V_3$.
We christen each such set $\left\{ V_1,\ V_2,\ V_3 \right\}$ a `structure'.
We have plotted the sixty structures that we have found
as sixty---blue, green, and red---points in figure~\ref{figura}.

We note that all the sixty structures that we have found would also
have been found if we had restricted our search to subgroups of $SU(3)$.
Namely, for all the structures,
there is always at least one case in
which both matrices $T_\ell$ and $T_\nu$ have determinant 1.
%%%%% NEW TEXT STARTS HERE.
We do not know why this happens but,
like Ludl has pointed out~\cite{ludl2},
it is possible that every finite subgroup of $U(3)$
is equivalent in physical terms to some finite subgroup of $SU(3)$,
because they can only produce the same Lagrangians.
%%%%% NEW TEXT ENDS HERE.

We divide all the structures into three types.
Each of these types is described in one of the following subsections.

\subsection{Structures on the blue curve}

The first type encompasses a total of 44 structures.
They are described by the analytical formula
\be
\label{suigp}
V_2 = \frac{1}{2} \left( 1 - V_1 - \sqrt{2 V_1 - 3 V_1^2} \right).
\ee
This is depicted as a blue curve in figure~\ref{figura}.
The relevance of that curve had already been noticed
in refs.~\cite{lim,grimuslavoura}.

All but two of the 44 structures on the blue curve may be written
\bs
\label{yutpe}
\ba
V_1 &=& \frac{1}{3}
\left( 1 - \cos{\frac{2 \pi k}{3 n}} \right),
\\
V_2 &=& \frac{1}{3}
\left[1 - \cos{\frac{2 \pi \left( k - n \right)}{3 n}} \right],
\\
V_3 &=& \frac{1}{3}
\left[1 - \cos{\frac{2 \pi \left( k + n \right)}{3 n}} \right],
\ea
\es
for positive integers $2 \le n \le 17$
and $k < n/2$ given in tables~\ref{tablenk}
\renewcommand{\arraystretch}{1.1}
\begin{table}[ht]
\centering
\begin{tabular}{|c||c|c|c|c|c|}
\hline
$n$ & 17 & 16 & 14 & 13 & 11
\\ \hline
$k$ & 1, 2, 3, 4, 5, 6, 7, 8 & 1, 3, 5, 7 & 1, 3, 5
& 1, 2, 3, 4, 5, 6 & 1, 2, 3, 4, 5
\\ \hline
\end{tabular}
\caption{The values of $n$ and $k$ to be used in equations~\eqref{yutpe}.}
\label{tablenk}
\end{table}
\renewcommand{\arraystretch}{1.0}
and~\ref{tablenk2}.
\vspace*{-5mm}
\renewcommand{\arraystretch}{1.1}
\begin{table}[ht]
\centering
\begin{tabular}{|c||c|c|c|c|c|c|c|c|c|}
\hline
$n$ & 10 & 9 & 8 & 7 & 6 & 5 & 4 & 3 & 2
\\ \hline
$k$ & 1, 3 & 1, 2, 4 & 1, 3 & 1, 2, 3 & 1 & 1, 2 & 1 & 1 & 1
\\ \hline
\end{tabular}
\caption{Continuation of table~\ref{tablenk}.}
\label{tablenk2}
\end{table}
\renewcommand{\arraystretch}{1.0}

\vspace*{5mm}

There are two more structures that are also described by equation~\eqref{suigp}
but cannot be described through equations~\eqref{yutpe}
for any small integers $n$ and $k$.
They are
\be
\label{ghoty}
\left\{ \frac{1}{ 6+\csc \frac{3\pi}{14} },\
\frac{1}{ 6-\sec \frac{\pi}{7} },\
\frac{1}{ 6-\csc \frac{\pi}{14} }
\right\}
\approx \left\{ 0.131512,\ 0.204495,\ 0.663993 \right\}
\ee
and
\be
\label{bhigo}
\left\{ \frac{3 - \sqrt{5}}{8},\ 1/4,\ \frac{3 + \sqrt{5}}{8} \right\}
\approx \left\{ 0.0954915,\ 0.25,\ 0.654508 \right\}.
\ee

The structures given by equations~\eqref{yutpe}
are produced by groups $\left\langle T_\ell,\ T_\nu \right\rangle$
%%%%% I HAVE MODIFIED THE FOLLOWING.
which are either $\Delta (6 n^2)$ or $D \left( 9n, 1, 1; 2, 1, 1 \right)$,
where $n$ is a positive integer~\cite{su3}.
We recall that those are subgroups of $SU(3)$.
The group $\Delta (6 n^2)$~\cite{6n2}
has $6 n^2$ elements and is generated by
\be
E = \left( \begin{array}{ccc} 0 & 1 & 0 \\ 0 & 0 & 1 \\ 1 & 0 & 0
\end{array} \right),
\quad
B = \left( \begin{array}{ccc} -1 & 0 & 0 \\ 0 & 0 & -1 \\ 0 & -1 & 0
\end{array} \right),
\quad \mbox{and} \quad
\left( \begin{array}{ccc} 1 & 0 & 0 \\ 0 & \eta & 0 \\ 0 & 0 & \eta^{-1}
\end{array} \right),
\ee
where $\eta = \exp{\left( 2 i \pi / n \right)}$.
The group $D \left( 9n, 1, 1; 2, 1, 1 \right)$
has $162 n^2$ elements and is generated by $E$,
$B$, and 
$\mathrm{diag} \left( \epsilon, \epsilon, \epsilon^{-2} \right)$,
where $\epsilon = \exp{\left[ 2 i \pi /
\! \left( 9 n \right) \right]}$~\cite{ludl1}.
In tables~\ref{Delta}
\renewcommand{\arraystretch}{1.3}
\begin{table}[h]
\centering
\begin{tabular}{|c|c|c|c|c|}
\hline
$\Delta \left( 6 \times 2^2 \right)$ & $\Delta \left( 6 \times 3^2 \right)$ &
$\Delta \left( 6 \times 4^2 \right)$ & $\Delta \left( 6 \times 5^2 \right)$ &
$\Delta \left( 6 \times 6^2 \right)$
\\*[0.0pt]
[24, 12] & [54, 8] & [96, 64] & [150, 5] & [216, 95]
\\ \hline
$\Delta \left( 6 \times 7^2 \right)$ & $\Delta \left( 6 \times 8^2 \right)$ &
$\Delta \left( 6 \times 9^2 \right)$ & $\Delta \left( 6 \times 10^2 \right)$ &
$\Delta \left( 6 \times 11^2 \right)$
\\*[0.0pt]
[294, 7] & [384, 568] & [486, 61] & [600, 179] & [726, 5]
\\ \hline
$\Delta \left( 6 \times 12^2 \right)$ & $\Delta \left( 6 \times 13^2 \right)$ &
$\Delta \left( 6 \times 14^2 \right)$ & $\Delta \left( 6 \times 15^2 \right)$ &
$\Delta \left( 6 \times 16^2 \right)$
\\*[0.0pt]
[864, 701] & [1014, 7] & [1176, 243] & [1350, 46] & [1536, 408544632]
\\ \hline
$\Delta \left( 6 \times 17^2 \right)$ & $\Delta \left( 6 \times 18^2 \right)$ &
 & &
\\*[0.0pt]
[1734, 5] & [1944, 849] & & &
\\ \hline
\end{tabular}
\caption{The {\tt SmallGroups} identifiers
of the groups $\Delta (6 n^2)$ with order smaller than 2\,000.}
\label{Delta}
\end{table}
\renewcommand{\arraystretch}{1.0}
and~\ref{D}
\renewcommand{\arraystretch}{1.1}
\begin{table}[h]
\centering
\begin{tabular}{|c|c|c|}
\hline
$n = 1$ & $n = 2$ & $n = 3$
\\*[0,0pt]
[162, 14] & [648, 259] & [1458, 659]
\\ \hline
\end{tabular}
\caption{The {\tt SmallGroups} identifiers of the groups
$D \left( 9 n, 1, 1; 2, 1, 1 \right) \equiv D^{(1)}_{9n, 3n}$~\cite{ludl1}
with order smaller than 2\,000.}
\label{D}
\end{table}
\renewcommand{\arraystretch}{1.0}
we give the {\tt SmallGroups} identifiers of these groups,
which have structures
\be
\left( \zz_n \times \zz_n \right) \rtimes S_3
\quad \mbox{and} \quad
\left( \zz_{9n} \times \zz_{3n} \right) \rtimes S_3,
\ee
respectively,
where $S_3$ is the permutation group of three objects.
% OLD TEXT
%described by {\tt GAP} as being of one of the following two forms:
%%
%\bs
%\ba
%\left[ \left( \zz_n \times \zz_n \right) \rtimes \zz_3 \right] \rtimes \zz_2
%& & \mbox{for}\ n\ \mathrm{not\ multiple\ of}\ 3,
%\label{delta} \\
%\left[ \left( \zz_{3n} \times \zz_n \right) \rtimes \zz_3 \right] \rtimes \zz_2
%& & \mbox{for}\ n\ \mathrm{multiple\ of}\ 3.
%\ea
%\es
%%
%All of these groups are subgroups of $SU(3)$
%(we have checked that
%their matrices $T_\ell$ and $T_\nu$ have both unit determinant).
%It can be shown that the groups in~\eqref{delta}
%are the well-known $\Delta (6 n^2)$.\footnote{We thank
%P.\ O.\ Ludl for demonstrating this for us.}

The structure~\eqref{ghoty} is produced in the main search
by the group $\left\langle T_\ell,\ T_\nu \right\rangle
%%%%% = [ 168, 42 ] = \mathrm{PSL} (3, 2)$.
= [168, 42] = \Sigma \left( 168 \right)$,
which is an `exceptional' finite subgroup of $SU(3)$.
The structure~\eqref{bhigo}
%%%%% is produced by
is produced by another exceptional subgroup of $SU(3)$,
the group
%%%%% $A_5$
$[60, 5] = \Sigma \left( 60 \right) \cong A_5$
(the symmetry group of the icosahedron)
in the secondary search~\cite{varzielas}.
That structure is the only one produced in the secondary search
that was not also a result of the main search;
it had already been found in refs.~\cite{varzielas,lavouraludl}.

It is convenient to number all the structures belonging to the blue curve
of figure~\ref{figura}
according to increasing values of $V_1$.
We thus construct the following list of the 44 structures:
\bs
\label{u8}
\ba
\mbox{structure 1}\ (n = 17,\ k = 1): & &
\left\{ 0.0025265,\ 0.463262,\ 0.534212 \right\};
\no
\mbox{structure 2}\ (n = 16,\ k = 1): & &
\left\{ 0.00285171,\ 0.460894,\ 0.536254 \right\};
\no
\mbox{structure 3}\ (n = 14,\ k = 1): & &
\left\{ 0.00372306,\ 0.455114,\ 0.541163 \right\};
\no
\mbox{structure 4}\ (n = 13,\ k = 1): & &
\left\{ 0.00431658,\ 0.451535,\ 0.544148 \right\};
\no
\mbox{structure 5}\ (n = 11,\ k = 1): & &
\left\{ 0.00602377,\ 0.442356,\ 0.55162 \right\};
\no
\mbox{structure 6}\ (n = 10,\ k = 1): & &
\left\{ 0.00728413,\ 0.436339,\ 0.556377 \right\};
\no
\mbox{structure 7}\ (n = 9,\ k = 1): & &
\left\{ 0.00898504,\ 0.428934,\ 0.562081 \right\};
\no
\mbox{structure 8}\ (n = 17,\ k = 2): & &
\left\{ 0.0100677,\ 0.424554,\ 0.565378 \right\};
\no
\mbox{structure 9}\ (n = 8,\ k = 1): & &
\left\{ 0.0113581,\ 0.419606,\ 0.569036 \right\};
\no
\mbox{structure 10}\ (n = 7,\ k = 1): & &
\left\{ 0.0148091,\ 0.407507,\ 0.577684 \right\};
\no
\mbox{structure 11}\ (n = 13,\ k = 2): & &
\left\{ 0.0171545,\ 0.400009,\ 0.582837 \right\};
\no
\mbox{structure 12}\ (n = 6,\ k = 1): & &
\left\{ 0.0201025,\ 0.391216,\ 0.588681 \right\};
\no
\mbox{structure 13}\ (n = 17,\ k = 3): & &
\left\{ 0.0225093,\ 0.384464,\ 0.593027 \right\};
\no
\mbox{structure 14}\ (n = 11,\ k = 2): & &
\left\{ 0.0238774,\ 0.380772,\ 0.595351 \right\};
\no
\mbox{structure 15}\ (n = 16,\ k = 3): & &
\left\{ 0.0253735,\ 0.376842,\ 0.597784 \right\};
\no
\mbox{structure 16}\ (n = 5,\ k = 1): & &
\left\{ 0.0288182,\ 0.368176,\ 0.603006 \right\};
\no
\mbox{structure 17}\ (n = 14,\ k = 3): & &
\left\{ 0.0330104,\ 0.358243,\ 0.608746 \right\};
\no
\mbox{structure 18}\ (n = 9,\ k = 2): & &
\left\{ 0.0354558,\ 0.352715,\ 0.611829 \right\};
\no
\mbox{structure 19}\ (n = 13,\ k = 3): & &
\left\{ 0.0381813,\ 0.346755,\ 0.615063 \right\};
\no
\mbox{structure 20}\ (n = 17,\ k = 4): & &
\left\{ 0.0396626,\ 0.343598,\ 0.616739 \right\};
\no
\mbox{structure 21}\ (n = 4,\ k = 1): & &
\left\{ 0.0446582,\ 0.333333,\ 0.622008 \right\};
\no
\mbox{structure 22}\ (n = 11,\ k = 3): & &
\left\{ 0.0529155,\ 0.317473,\ 0.629612 \right\};
\no
\mbox{structure 23}\ (n = 7,\ k = 2): & &
\left\{ 0.0579204,\ 0.308423,\ 0.633656 \right\};
\no
\mbox{structure 24}\ (n = 17,\ k = 5): & &
\left\{ 0.0612677,\ 0.302577,\ 0.636155 \right\};
\no
\mbox{structure 25}\ (n = 10,\ k = 3): & &
\left\{ 0.063661,\ 0.298491,\ 0.637848 \right\};
\no
\mbox{structure 26}\ (n = 13,\ k = 4): & &
\left\{ 0.0668524,\ 0.293154,\ 0.639993 \right\}; \hspace*{7mm}
\no
\mbox{structure 27}\ (n = 16,\ k = 5): & &
\left\{ 0.0688822,\ 0.289825,\ 0.641293 \right\};
\no
\mbox{structure 28}\ (n = 3,\ k = 1): & &
\left\{ 0.0779852,\ 0.275451,\ 0.646564 \right\}; \hspace*{7mm}
\no
\mbox{structure 29}\ (n = 17,\ k = 6): & &
\left\{ 0.086997,\ 0.262022,\ 0.650981 \right\};
\no
\mbox{structure 30}\ (n = 14,\ k = 5): & &
\left\{ 0.0889827,\ 0.25916,\ 0.651858 \right\};
\no
\mbox{structure 31}\ (n = 11,\ k = 4): & &
\left\{ 0.0920887,\ 0.254747,\ 0.653164 \right\};
\no
\mbox{structure 32}: & & \left\{ 0.0954915,\ 0.25,\ 0.654508 \right\};
\no
\mbox{structure 33}\ (n = 8,\ k = 3): & &
\left\{ 0.0976311,\ 0.24706,\ 0.655309 \right\};
\no
\mbox{structure 34}\ (n = 13,\ k = 5): & &
\left\{ 0.102425,\ 0.240594,\ 0.656981 \right\};
\no
\mbox{structure 35}\ (n = 5,\ k = 2): & &
\left\{ 0.11029,\ 0.230328,\ 0.659383 \right\};
\no
\mbox{structure 36}\ (n = 17,\ k = 7): & &
\left\{ 0.116461,\ 0.222548,\ 0.660991 \right\};
\no
\mbox{structure 37}\ (n = 7,\ k = 3): & &
\left\{ 0.125503,\ 0.211553,\ 0.662944 \right\};
\no
\mbox{structure 38}\ (n = 16,\ k = 7): & &
\left\{ 0.130413,\ 0.205772,\ 0.663815 \right\};
\no
\mbox{structure 39}: & & \left\{ 0.131512,\ 0.204495,\ 0.663993 \right\};
\no
\mbox{structure 40}\ (n = 9,\ k = 4): & &
\left\{ 0.13428,\ 0.201307,\ 0.664413 \right\};
\no
\mbox{structure 41}\ (n = 11,\ k = 5): & &
\left\{ 0.139981,\ 0.194862,\ 0.665157 \right\};
\no
\mbox{structure 42}\ (n = 13,\ k = 6): & &
\left\{ 0.143978,\ 0.190436,\ 0.665586 \right\};
\no
\mbox{structure 43}\ (n = 17,\ k = 8): & &
\left\{ 0.149212,\ 0.184754,\ 0.666034 \right\};
\no
\mbox{structure 44}\ (n = 2,\ k = 1): & &
\left\{ 1/6,\ 1/6,\ 2/3 \right\}.
\nonumber
\ea
\es

\subsection{Structures on the green line}

The second type of structures features two equal $V_j$,
\textit{i.e.}\ either $V_1 = V_2$ or $V_2 = V_3$.
These structures straddle the green line in figure~\ref{figura}.
One of them is of course structure~44,
which corresponds to the point where the green line meets the blue curve.
There are six more such structures on the green line.
They are
\bs
\ba
\mbox{structure 45}\
\left( V_1 = \frac{3 - \sqrt{5}}{12} \right): & &
\left\{ 0.063661,\ 0.063661,\ 0.872678 \right\};
\no
\mbox{structure 46}\
\left( V_1 = \frac{5 - \sqrt{5}}{20} \right): & &
\left\{ 0.138197,\ 0.138197,\ 0.723607 \right\};
\no
\mbox{structure 47}\
\left( V_1 = \frac{1}{4} \right): & &
\left\{ 0.25,\ 0.25,\ 0.5 \right\};
\no
\mbox{structure 48}\
\left( V_1 = \frac{1}{3} \right): & &
\left\{ 0.333333,\ 0.333333,\ 0.333333 \right\};
\no
\mbox{structure 49}\
\left( V_1 = \frac{5 - \sqrt{5}}{10} \right): & &
\left\{ 0.276393,\ 0.361803,\ 0.361803 \right\};
\no
\mbox{structure 50}\
\left( V_1 = \frac{3 - \sqrt{5}}{6} \right): & &
\left\{ 0.127322,\ 0.436339,\ 0.436339 \right\}.
\nonumber
\ea
\es

The structures 45, 46, 49, and 50 all originate
in $\left\langle T_\ell,\ T_\nu \right\rangle = A_5$.
%%%%% which has {\tt GAP} identifier [60, 5].
Structure~47 originates in the permutation group
%%%%% $S_4$,
$S_4 \cong \Delta (6 \times 2^2)$.
%%%%% with {\tt GAP} identifier [24, 12].
Structure~48 originates in the alternating group
%%%%% $A_4$,
$A_4 \cong \Delta (3 \times 2^2)$,
%%%%% with {\tt GAP} identifier [12, 3].
with {\tt SmallGroups} identifier [12, 3].

Both structures~46 and~49 justify the \textit{Ansatz}~\cite{goldenratio}
\be
\cot{\theta_{12}} = \varphi \equiv \frac{1 + \sqrt{5}}{2},
\ee
relating the lepton mixing angle $\theta_{12}$ to the `golden ratio' $\varphi$.
Indeed,
$2 V_1 =
1 \! \left/ \left( 1 + \varphi^2 \right) \right.$ in structure~46
%%%%% and { $ V_1 =
and $ V_1 =
1 \! \left/ \left( 1 + \varphi^2 \right) \right.$ in structure~49.

\subsection{Isolated structures}

Besides the points on the blue curve and the points on the green line,
there are ten isolated points marked red in figure~\ref{figura}.
The corresponding structures are
\ba
& & \mbox{structure~51}:\
\left\{0.0347854,\ 0.166667,\ 0.798548 \right\}
\no*[1mm]
&=& \left\{ \frac{5 - \sqrt{21}}{12},\ \frac{1}{6},\ \frac{5 + \sqrt{21}}{12}
\right\};
\no*[2mm]
& & \mbox{structure~52}:\
\left\{ 0.0389375,\ 0.306554,\ 0.654508 \right\}
\no*[1mm]
&=& \left\{ \frac{5 + \sqrt{3} - \sqrt{5} - \sqrt{15}}{16},\
\frac{5 - \sqrt{3} - \sqrt{5} + \sqrt{15}}{16},\
\frac{3 + \sqrt{5}}{8}
\right\};
\no*[2mm]
& & \mbox{structure~53}:\
\left\{ 0.0442811,\ 0.25,\ 0.705719 \right\} =
\left\{ \frac{3 - \sqrt{7}}{8},\ \frac{1}{4},\ \frac{3 + \sqrt{7}}{8}
\right\};
\no*[2mm]
& & \mbox{structure~54}:\
\left\{ 0.0625591,\ 0.138197,\ 0.799244 \right\}
\no*[1mm]
&=& \left\{ \frac{15 + \sqrt{5} - \sqrt{150 + 30 \sqrt{5}}}{40},\
\frac{5 - \sqrt{5}}{20},\
\frac{15 + \sqrt{5} + \sqrt{150 + 30 \sqrt{5}}}{40} \right\};
\no*[2mm]
& & \mbox{structure~55}:\
\left\{ 0.08592426701,\ 0.201689718788,\ 0.712386014201 \right\}
\no*[1mm]
&=& \left\{ \frac{2}{3 \left( 2+\csc \frac{\pi}{18} \right)},\
\frac{2}{3 \left( 2+\sec \frac{2 \pi}{9} \right)},\
\frac{2}{3 \left( 2-\sec \frac{\pi}{9} \right)}
\right\};
\no*[2mm]
& & \mbox{structure~56}:\
\left\{ 0.0914501,\ 0.361803,\ 0.546747 \right\}
\no*[1mm]
&=& \left\{ \frac{15 - \sqrt{5} - \sqrt{150 - 30 \sqrt{5}}}{40},\
\frac{5 + \sqrt{5}}{20},\
\frac{15 - \sqrt{5} + \sqrt{150 - 30 \sqrt{5}}}{40} \right\};
\no*[2mm]
& & \mbox{structure~57}:\
\left\{ 0.0954915,\ 0.101940,\ 0.802569 \right\}
\no*[1mm]
&=& \left\{ \frac{3 - \sqrt{5}}{8},
\frac{5 - \sqrt{3} + \sqrt{5} - \sqrt{15}}{16},\
\frac{5 + \sqrt{3} + \sqrt{5} + \sqrt{15}}{16}
\right\};
\no*[2mm]
& & \mbox{structure~58}:\
\left\{ 0.105662,\ 0.394338,\ 0.5 \right\} =
\left\{ \frac{3 - \sqrt{3}}{12},\ \frac{3 + \sqrt{3}}{12},\ \frac{1}{2}
\right\};
\no*[2mm]
& & \mbox{structure~59}:\
\left\{ 0.158494,\ 0.25,\ 0.591506 \right\} =
\left\{ \frac{3 - \sqrt{3}}{8},\ \frac{1}{4},\ \frac{3 + \sqrt{3}}{8}
\right\};
\no*[2mm]
& & \mbox{structure~60}:\
\left\{ 1/6,\ 1/3,\ 1/2 \right\}.
\nonumber
\ea

All structures~51--60 but structures~55 and~58
are either rows or columns of matrices found in ref.~\cite{fonseca};
we have taken their analytic expressions from that paper.

Structures~51--60 originate in the following
groups $\left\langle T_\ell,\ T_\nu \right\rangle$:
%%%%% I HAVE REWRITTEN THE FOLLOWING ``ITEMIZE'' SECTION.
%
\begin{itemize}
\item The exceptional $SU(3)$ subgroup $\Sigma \left( 168 \right)$,
which has {\tt SmallGroups} identifier [168, 42],
for structures~51 and~53;
\item The exceptional $SU(3)$ subgroup
$\Sigma \left( 360 \times 3 \right)$,
which has {\tt SmallGroups} identifier [1080, 260],
for structures~52, 54, 56, and~57;
\item any of three groups
of order 648---with {\tt SmallGroups} identifiers [648, 531],
[648, 532],
and [648, 533]---for structures~55,
58,
and~60.
Those three groups have similar structure
$\left\{ \left[ \left( \zz_3 \times \zz_3 \right) \rtimes \zz_3 \right]
\rtimes Q_8 \right\} \rtimes \zz_3$,
where $Q_8$ is the quaternion group,
a subgroup of $SU(2)$.
The group [648, 532] is the $SU(3)$ exceptional subgroup
$\Sigma \left( 216 \times 3 \right)$.
\item for structure~59,
$\left\langle T_\ell,\ T_\nu \right\rangle$ may be either
the exceptional $SU(3)$ subgroup $\Sigma \left( 36 \times 3 \right)$,
with {\tt SmallGroups} identifier [108, 15],
or some other groups with analogous structures
and orders which are multiple of 108,
like [216, 25], [324, 111], [432, 57], and so on.
\end{itemize}
%
%%%%% I HAVE ADDED
Table~\ref{exceptional} gives the {\tt SmallGroups} identifiers
of all six exceptional subgroups of $SU(3)$.
\renewcommand{\arraystretch}{1.1}
\begin{table}[h]
\centering
\begin{tabular}{|c|c|c|c|c|c|}
\hline
$\Sigma \left( 60 \right)$ & $\Sigma \left( 168 \right)$ &
$\Sigma \left( 36 \times 3 \right)$ & $\Sigma \left( 72 \times 3 \right)$ &
$\Sigma \left( 216 \times 3 \right)$ & $\Sigma \left( 360 \times 3 \right)$
\\*[0.0pt]
[60, 5] & [168, 42] & [108, 15] & [216, 88] & [648, 532] & [1080, 260]
\\ \hline
\end{tabular}
\caption{The {\tt SmallGroups} identifiers
of the exceptional finite subgroups of $SU(3)$.}
\label{exceptional}
\end{table}
\renewcommand{\arraystretch}{1.0}

\section{Comparison with the data} \label{comparison}

The matrix $\left| U \right|^2$ is parameterized as
\be
\left| U \right|^2 = \left( \begin{array}{ccc}
c_{12}^2 c_{13}^2 & s_{12}^2 c_{13}^2 & s_{13}^2 \\
s_{12}^2 c_{23}^2 + c_{12}^2 s_{23}^2 s_{13}^2 + Y &
c_{12}^2 c_{23}^2 + s_{12}^2 s_{23}^2 s_{13}^2 - Y & s_{23}^2 c_{13}^2 \\
s_{12}^2 s_{23}^2 + c_{12}^2 c_{23}^2 s_{13}^2 - Y &
c_{12}^2 s_{23}^2 + s_{12}^2 c_{23}^2 s_{13}^2 + Y & c_{23}^2 c_{13}^2
\end{array} \right),
\ee
where $c_{ij} \equiv \cos{\theta_{ij}}$ and $s_{ij} \equiv \sin{\theta_{ij}}$
for $(ij) = (12), (23), (13)$.
The quantity $Y \equiv 2 c_{12} s_{12} c_{23} s_{23} s_{13} \cos{\delta}$.

There are in the literature three global phenomenological fits
to the parameters $\theta_{12}$,
$\theta_{23}$,
$\theta_{13}$,
and $\delta$.
The specific bounds on each parameter depend on which of refs.~\cite{tortola},
\cite{fogli}, or~\cite{schwetz} one uses
and also on whether a `normal' or `inverted' ordering
is assumed for the neutrino masses.
For definiteness,
we shall use the values of ref.~\cite{tortola} for an inverted ordering.
They are
\be
s_{12}^2 \in \left[ 0.278,\ 0.375 \right], \quad
s_{23}^2 \in \left[ 0.403,\ 0.640 \right], \quad
s_{13}^2 \in \left[ 0.0183,\ 0.0297 \right]
\ee
at $3 \sigma$ level,
\be
s_{12}^2 \in \left[ 0.292,\ 0.357 \right], \quad
s_{23}^2 \in \left[ 0.432,\ 0.621 \right], \quad
s_{13}^2 \in \left[ 0.0202,\ 0.0278 \right]
\ee
at $2 \sigma$ level,
and
\be
\begin{array}{lcl}
s_{12}^2 \in \left[ 0.307,\ 0.339 \right],
& &
s_{23}^2 \in \left[ 0.530,\ 0.598 \right],
\\*[1mm]
s_{13}^2 \in \left[ 0.0221,\ 0.0259 \right],
& &
\delta \in \left[ 1.16\, \pi,\ 1.82\, \pi \right]
\end{array}
\ee
at $1 \sigma$ level.
Note that $\cos{\delta}$ is free at both $3 \sigma$ and $2 \sigma$ levels.

Up to now,
we have considered that the structures correspond to predictions
for the third row of $\left| U \right|^2$.
This is because $T_\ell$ and $H_\ell$ are simultaneously diagonalizable
and we have assumed that the eigenvalues of $T_\ell$ obey $l_1 = l_2 \neq l_3$.
However,
\begin{itemize}
\item we might instead have assumed either $l_1 = l_3 \neq l_2$
or $l_2 = l_3 \neq l_1$,
and then we would have predicted,
through exactly the same mathematics,
either the second or the first row,
respectively,
of $\left| U \right|^2$;
\item equations~\eqref{sdihp} are invariant under simultaneous permutations
of
\be
\left( \left| U_{31} \right|^2,\ \left| U_{32} \right|^2,\
\left| U_{33} \right|^2 \right)
\quad \mbox{and} \quad \left( n_1,\ n_2,\ n_3 \right),
\ee
hence,
by altering the ordering of the eigenvalues of $T_\nu$
one may alter the ordering of the matrix elements
in the row of $\left| U \right|^2$ that one is predicting;
\item $\left( H_\ell,\ T_\ell \right)$ and $\left( H_\nu,\ T_\nu \right)$
may be interchanged in the reasoning of section~\ref{theory},
and then we would be predicting a \emph{column}\/ instead of a \emph{row}\/
of $\left| U \right|^2$.
\end{itemize}
Thus,
the structures in the previous
section must be interpreted as predictions for either \emph{any row}\/
or \emph{any column}\/ of $\left| U \right|^2$;
moreover,
for each such column or row of $\left| U \right|^2$
the three numbers of the structure may be taken in any order.

We may now state the results of the confrontation of our structures
with the data:
\begin{description}
\item If one tries to fit the first row of $\left| U \right|^2$
through one of our structures,
one finds that none of them is able to do it.
Indeed,
though structures~12--16 have $V_1$ adequate
to fit $\left| U_{13} \right|^2$,
their $V_2$ is much too large
to agree with the upper bound on $s_{12}^2$.
\item If one tries to fit the third column of $\left| U \right|^2$
through one of our structures,
one finds that structures~12 and~16 do it at $3\sigma$ level
and structures~13--15 manage it at $2\sigma$ level.
\item Many structures on the blue curve of figure~\ref{figura}
fit the first column of $\left| U \right|^2$:
structures~29--44 do it at $1\sigma$ level,
structures~26--28 at $2\sigma$ level,
and structures~24 and~25 at $3\sigma$ level.
The structures on the green line in figure~1
(with the exception of structure~44)
and the isolated structures in that figure
cannot fit the first column of $\left| U \right|^2$,
even though some of the corresponding points appear close to the blue
line in figure~\ref{figura}.
Notice that structure~32 incorporates
one of the `golden ratio' \textit{Ans\"atze}\/
for the mixing angle $\theta_{12}$,
namely $\cos{\theta_{12}} = \varphi / 2\, \Leftrightarrow\,
\cos^2{\theta_{12}}
= \left. \left( 3 + \sqrt{5} \right) \right/ \! 8$~\cite{rodejohann};
it was already pointed out in ref.~\cite{varzielas} that that particular
structure gives an excellent fit to the first column of $\left| U \right|^2$.
\item The second column of $\left| U \right|^2$
is fitted at $2 \sigma$ level by structure~48
and at $3 \sigma$ level by structures~49 and~60.
Structure~48 corresponds
to the well-known `trimaximal mixing'~\cite{trimaximal}
or `TM$_2$'~\cite{TM2}
\textit{Ansatz}\/ for that column of $\left| U \right|^2$.
Structure~49 corresponds to one of the two~\cite{adulpravitchai}
`golden ratio' \textit{Ans\"atze}\/ for $\left| U_{12} \right|^2$.
\item The second row of $\left| U \right|^2$ may be fitted
at $1\sigma$ level by structure~56,
at $2\sigma$ level by structures~47, 50, and~58--60,
and at $3\sigma$ level by structure~21.
\item The third row of $\left| U \right|^2$ may be fitted
at $1\sigma$ level by structure~50,
at $2\sigma$ level by structures~47, 56, 58, and~60,
and at $3\sigma$ level by structure~49.
The adequateness of structure~47 had already been pointed out
in ref.~\cite{lavouraludl}.
\end{description}

It should be stressed that the fits in this section
are rather sensitive to the precise phenomenlogical bounds that one utilizes
for the three $s_{ij}^2$ and for $\cos{\delta}$.
We have used here the bounds in ref.~\cite{tortola}
with an inverted neutrino mass ordering but,
if we had instead used the bounds
in either ref.~\cite{fogli} or ref.~\cite{schwetz},
or the bounds for a normal ordering,
then our results would have differed somewhat.

\section{Summary} \label{conclusions}

In this paper we have assumed that neutrinos are Dirac particles,
\textit{i.e.}\ that the lepton sector is similar to the quark sector.
We have assumed that mixing in the lepton sector originates
from a discrete horizonal symmetry group $G$
that breaks into two distinct subgroups $G_\ell$ and $G_\nu$
under which the mass matrices of the charged leptons and of the neutrinos
are separately invariant.
We have considered the special situation in which one of the subgroups
$G_{\ell,\nu}$ is generated by a matrix with two equal eigenvalues
while the other subgroup
is generated by a matrix with non-degenerate eigenvalues.
Under these assumptions,
by making a {\tt GAP/SmallGroups} search
of all possible groups $G$ of order smaller than 2\,000,
we have found 60 possible structures for either a row or a column
of the lepton mixing matrix.
Several of those structures constitute realistic
predictions for either any of the columnsor for the second row
or for the third row of the PMNS matrix.

%%%%% I HAVE WRITTEN THIS NEW APPENDIX.
\begin{appendix}

\section{Nilpotent groups and groups with a normal Sylow 3-subgroup}
\label{groups}

\setcounter{equation}{0}
\renewcommand{\theequation}{A\arabic{equation}}

\subsection{Nilpotent groups}

The \emph{commutator}\/ $\left[ f, g \right]$
of two group elements $f$ and $g$ is the group element $f^{-1} g^{-1 } f g$.
Clearly,
if $f$ commutes with $g$,
\textit{i.e.}\ if $f g = g f$,
then $\left[ f, g \right]$ is the identity element $e$ of the group;
the converse is also true:
if $\left[ f, g \right] = e$ then $f g = g f$.

Let $G$ be a finite group.
Let $F$ be one of its subgroups.
Then,
we define $\left[ F, G \right]$ as the subgroup of $G$
generated by all the elements $\left[ f, g \right]$ of $G$
which are the commutators of some $f \in F$ and some $g \in G$.
(Notice that the commutators do not in general close
under the group multiplication;
therefore,
one must use them to \emph{generate}\/ a subgroup of $G$.)

We next define the \emph{descending central series}\/ of a group $G$.
This is the series $G, G_1, G_2, G_3, \ldots$ of subgroups of $G$
defined through the procedure $G_1 = \left[ G, G \right]$,
$G_2 = \left[ G_1, G \right]$,
$G_3 = \left[ G_2, G \right]$,
and so on.

We finally define a \emph{nilpotent}\/ group:
it is a group whose descending central series ends up
in the trivial group---the one consisting solely of the identity element.

Let us give an example of a nilpotent group:
the $D_4$ group generated by the two matrices
\be
\label{dd44}
A = \left( \begin{array}{cc} 0 & 1 \\ 1 & 0 \end{array} \right)
\quad \mbox{and} \quad
B = \left( \begin{array}{cc} 1 & 0 \\ 0 & -1 \end{array} \right).
\ee
Clearly,
\be
\label{uhiy}
\left[ A, A \right] = \left[ B, B \right] =
\left( \begin{array}{cc} 1 & 0 \\ 0 & 1 \end{array} \right)
\quad \mbox{and} \quad
\left[ A, B \right] =
\left( \begin{array}{cc} -1 & 0 \\ 0 & -1 \end{array} \right).
\ee
Therefore,
$\left[ D_4, D_4 \right] = \mathbbm{Z}_2$
is formed by the two matrices in~\eqref{uhiy}.
Since those two matrices commute with all the matrices of $D_4$
(indeed, they commute with any $2 \times 2$ matrix),
$\left[ \left[ D_4, D_4 \right], D_4 \right]$
is just the unit $2 \times 2$ matrix,
\textit{i.e}\ it is the trivial group.
Therefore,
$D_4$ is nilpotent.

We next give an example of a non-nilpotent group:
the $D_3$ group generated by the two matrices
\be
\label{ss33}
A = \left( \begin{array}{cc} 0 & 1 \\ 1 & 0 \end{array} \right)
\quad \mbox{and} \quad
C = \left( \begin{array}{cc} \omega & 0 \\ 0 & \omega^2 \end{array} \right),
\ee
where $\omega = \exp{\left( 2 i \pi / 3 \right)}$.
Clearly,
\be
\label{upiy}
\left[ A, A \right] = \left[ C, C \right] =
\left( \begin{array}{cc} 1 & 0 \\ 0 & 1 \end{array} \right)
\quad \mbox{and} \quad
\left[ A, C \right] =
\left( \begin{array}{cc} \omega^2 & 0 \\ 0 & \omega \end{array} \right) = C^{-1}.
\ee
Therefore,
$\left[ D_3, D_3 \right] = \mathbbm{Z}_3$
is the group generated by $C^{-1}$.
We compute $\left[ C^{-1}, A \right] = C^{-1}$
to conclude that $\left[ \left[ D_3, D_3 \right], D_3 \right]$
is again $\mathbbm{Z}_3$.
Therefore,
the descending central series of $D_3$ is $D_3$,
$\mathbbm{Z}_3$,
$\mathbbm{Z}_3$,
$\mathbbm{Z}_3, \ldots$;
this central series never ends up in the trivial group,
hence $D_3$ is not nilpotent.

\subsection{Groups with a normal Sylow 3-subgroup}

Let $G$ be a finite group.
Let $F$ be one of its subgroups.
Let $g \in G$ be an element of $G$.
Then,
the \emph{left coset}\/ of $F$ with respect to $g$
is the set of all elements $h \in G$ that may be written $h = g f$
for some $f \in F$.
Similarly,
the \emph{right coset}\/ of $F$ with respect to $g$
is defined to be the set of all the $h^\prime \in G$
that may be written $h^\prime = f g$ for some $f \in F$.
The subgroup $F$ of $G$ is said to be \emph{normal}\/
if its left coset with respect to any $g \in G$
coincides with the right coset with respect to $g$.

Take for instance the group $D_3$ generated by the matrices in~\eqref{ss33}.
It has a $\mathbbm{Z}_3$ subgroup formed by $C$,
$C^2$,
and the $2 \times 2$ unit matrix.
It is easily seen that $A C = C^2 A$ and $A C^2 = C A$.
Therefore,
the left and right cosets of $\mathbbm{Z}_3$ relative to $A$ are identical;
hence,
$\mathbbm{Z}_3$ is a normal subgroup of $D_3$.

The \emph{order}\/ of an element $g$ of a finite group $G$
is the smallest positive integer $o$
such that $g^o = e$ is the identity of $G$.

Let $p$ be a prime number,
then a $p$-group is a group where \emph{all}\/ the elements have order $o$
which is a power of $p$
(different elements may have different orders,
but all the orders are powers of $p$).
Thus,
a 3-group is a group where all the elements either have order one,
or order three,
or order nine,
\textit{etc.}
An obvious example is the well-known group $\Delta(27)$,
generated by
\be
\left( \begin{array}{ccc} 0 & 1 & 0 \\ 0 & 0 & 1 \\ 1 & 0 & 0
\end{array} \right)
\quad \mbox{and} \quad
\left( \begin{array}{ccc} 1 & 0 & 0 \\ 0 & \omega & 0 \\ 0 & 0 & \omega^2
\end{array} \right);
\ee
all 27 elements of $\Delta(27)$ except the identity have order three.
Similarly,
the group $D_4$ generated by the matrices in~\eqref{dd44} is a 2-group---all
its elements either have order one, or order two, or order four.
In general,
\emph{a finite group is a $p$-group if and only if
the number of elements of the group is a power of $p$};
thus,
the 3-groups are the groups with $3^n$ elements,
for some integer $n$.

A subgroup $S$ of a group $G$
is called a \emph{Sylow $p$-subgroup}\/ if it is a $p$-group
and if there is no larger $p$-subgroup of $G$ that contains $S$
as a proper subgroup.

Thus,
a normal Sylow 3-subgroup $S$ of a finite group $G$
is a normal subgroup of $G$ with $3^n$ elements
such that there is no subgroup of $G$ with $3^m$ elements,
$m > n$,
that contains $S$.

For instance,
the $\mathbbm{Z}_3$ subgroup of $D_3$
generated by the matrix $C^{-1}$ in~\eqref{upiy} is a Sylow 3-subgroup of $D_3$;
indeed,
the group $D_3$ has six elements and therefore any three-element subgroup of it
is necessarily a Sylow 3-subgroup.
Since we already know that $\mathbbm{Z}_3$ is a normal subgroup of $D_3$,
we conclude that $\mathbbm{Z}_3$ is a normal Sylow 3-subgroup of $D_3$.
The conjecture mentioned in section~\ref{searches} then informs us that
$D_3$ does \emph{not}\/ have a faithful three-dimensional irrep;
this is indeed true.

In the same way,
a normal Sylow 3-subgroup of a group $G$ of order $1536 = 3 \times 2^9$
is just a normal $\mathbbm{Z}_3$ subgroup of $G$.

A different example is the group $A_4$ generated by
\be
\label{uoiyp}
D = \left( \begin{array}{ccc} 1 & 0 & 0 \\ 0 & -1 & 0 \\ 0 & 0 & -1
\end{array} \right)
\quad \mbox{and} \quad
E = \left( \begin{array}{ccc} 0 & 1 & 0 \\ 0 & 0 & 1 \\ 1 & 0 & 0
\end{array} \right).
\ee
The group $A_4$ has $12 = 3 \times 2^2$ elements
and therefore its $\mathbbm{Z}_3$ subgroup formed by $E$,
$E^2$,
and the unit $3 \times 3$ matrix is automatically a Sylow 3-subgroup.
However,
it is not a normal subgroup,
because the set $\left\{ D E, D E^2 \right\}$
does not coincide with the set $\left\{ E D, E^2 D \right\}$.
Thus,
$A_4$ does not have a normal Sylow-3 subgroup
and it is allowed by the conjecture of section~\ref{searches}
to have a faithful three-dimensional irrep;
this is indeed the representation generated by the matrices in~\eqref{uoiyp}.

\end{appendix}

\paragraph{Acknowledgements:}
We gratefully acknowledge the collaboration of Pa\-trick Otto Ludl
in the beginning of this work;
we also thank him for valuable discussions all along.
The work of D.J.\ was supported by the Lithuanian Academy of Sciences
through project DaFi2016.
The work of L.L.\ was supported by
the Portuguese \textit{Funda\c{c}\~ao para a Ci\^encia e a
Tecnologia} through the projects CERN/FIS-NUC/0010/2015
and UID/FIS/00777/2013,
which are partially funded by POCTI (FEDER),
COMPETE,
QREN,
and the European Union.

%%%%% \begin{thebibliography}{9}


\begin{thebibliography}{99}

\bibitem{lam}
C.~S.~Lam,
\textit{Determining Horizontal Symmetry from Neutrino Mixing},
Phys.\ Rev.\ Lett.\ {\bf 101} (2008) 121602
[{\tt arXiv:0804.2622 [hep-ph]}];
\\
C.~S.~Lam,
\textit{Unique horizontal symmetry of leptons},
Phys.\ Rev.\ D {\bf 78} (2008) 073015
[{\tt arXiv:0809.1185 [hep-ph]}].

\bibitem{hagedorn}
R.~de Adelhart Toorop, F.~Feruglio, and C.~Hagedorn,
\textit{Finite modular groups and lepton mixing},
Nucl.\ Phys.\ B {\bf 858} (2012) 437
[{\tt arXiv: 1112.1340 [hep-ph]}];
\\
C.~Hagedorn, A.~Meroni, and L.~Vitale,
\textit{Mixing patterns from the groups $\Sigma \left( n \varphi \right)$},
J.\ Phys.\ A {\bf 47} (2014) 055201
[{\tt arXiv:1307.5308 [hep-ph]}].

\bibitem{GAP}
\textit{GAP---Groups, Algorithms, Programming---A system for
computational discrete algebra},
{\tt http://www.gap-system.org/}.

\bibitem{SG}
H.\ U.\ Besche, B.\ Eick, and E.\ A.\ O'Brien,
\textit{A millenium project: constructing Small Groups},
Intern.\ J.\ Algebra Comput.\ {\bf 12} (2002) 623;
\\
H.\ U.\ Besche, B.\ Eick, and E.\ A.\ O'Brien,
\textit{The SmallGroups library},
{\tt http://www.gap-system.org/Packages/sgl.html}
and {\tt http:// www.icm.tu-bs.de/ag\_algebra/software/small/}.

\bibitem{lam2}
C.~S.~Lam,
\textit{Finite symmetry of leptonic mass matrices},
Phys.\ Rev.\ D {\bf 87} (2013) 013001
[{\tt arXiv:1208.5527 [hep-ph]}].

\bibitem{lim}
M.~Holthausen, K.~S.~Lim, and M.~Lindner,
\textit{Lepton mixing patterns from a scan of finite discrete groups},
Phys.\ Lett.\ B {\bf 721} (2013) 61
[{\tt arXiv:1212.2411 [hep-ph]}].

\bibitem{fonseca}
R.~M.~Fonseca and W.~Grimus,
\textit{Classification of lepton mixing matrices
from finite residual symmetries},
J.\ High Energy Phys.\ {\bf 1409} (2014) 033
[{\tt arXiv:1405.3678 [hep-ph]}].

\bibitem{quarks}
T.~Araki, H.~Ishida, H.~Ishimori, T.~Kobayashi, and A.~Ogasahara,
\textit{CKM matrix and flavor symmetries},
Phys.\ Rev.\ D {\bf 88} (2013) 096002
[{\tt arXiv:1309.4217 [hep-ph]}].

\bibitem{chinese}
C.-Y.~Yao and G.-J.~Ding,
\textit{Lepton and quark mixing patterns from finite flavor symmetries},
Phys.\ Rev.\ D {\bf 92} (2015) 096010
[{\tt arXiv:1505.03798 [hep-ph]}].

\bibitem{lavouraludl}
L.~Lavoura and P.~O.~Ludl,
\textit{Residual $\mathbb{Z}_2 \times \mathbb{Z}_2$ symmetries
and lepton mixing},
Phys.\ Lett.\ B {\bf 731} (2014) 331
[{\tt arXiv:1401.5036 [hep-ph]}].

\bibitem{Branco}
G.~C.~Branco and L.~Lavoura,
\textit{Rephasing-invariant parametrization of the quark mixing matrix},
Phys.\ Lett.\ B {\bf 208} (1988) 123.

\bibitem{radovcic}
W.~Grimus, L.~Lavoura, and B.~Radov\v{c}i\'c,
\textit{Type II seesaw mechanism for Higgs doublets
and the scale of new physics},
Phys.\ Lett.\ B {\bf 674} (2009) 117
[{\tt arXiv:0902.2325 [hep-ph]}].

\bibitem{valle}
See for instance M.~Reig, J.~W.~F.~Valle, and C.~A.~Vaquera-Araujo,
\textit{Realistic $\mathrm{SU(3)_c \otimes SU(3)_L \otimes U(1)_X}$ model
with type-II Dirac neutrino seesaw mechanism},
{\tt arXiv:1606.08499 [hep-ph]},
and the references therein.

%%%%% NEW REFERENCES
\bibitem{talbert}
J.~Talbert,
\textit{[Re]constructing finite flavour groups:
horizontal symmetry scans from the bottom-up},
J.\ High Energy Phys.\ {\bf 1412} (2014) 058
[{\tt arXiv:1409.7310 [hep-ph]}];
\\
I.~de Medeiros Varzielas, R.~W.~Rasmussen, and J.~Talbert,
\textit{Bottom-up discrete symmetries for Cabibbo mixing},
{\tt arXiv:1605.03581 [hep-ph]}.

\bibitem{grouptheory}
See for instance I.~M.~Isaacs,
\textit{Finite group theory}
[American Mathematical Society, 2008].
%%%%%

\bibitem{kingludl}
S.~F.~King and P.~O.~Ludl,
\textit{Direct and semi-direct approaches
to lepton mixing with a massless neutrino},
J.\ High Energy Phys.\ {\bf 1606} (2016) 147
[{\tt arXiv:1605.01683 [hep-ph]}].

%%%%% NEW REFERENCE
\bibitem{ludl2}
P.~O.~Ludl,
\textit{On the finite subgroups of $U(3)$ of order smaller than 512},
J.\ Phys.\ A {\bf 43} (2010) 395204
[Erratum: \textit{ibid.}\ {\bf 44} (2011) 139501]
[{\tt arXiv:1006.1479 [math-ph]}].
%%%%% END OF NEW REFERENCE


\bibitem{grimuslavoura}
W.~Grimus and L.~Lavoura,
\textit{Double seesaw mechanism and lepton mixing},
J.\ High Energy Phys.\ {\bf 1403} (2014) 004
[{\tt arXiv:1309.3186 [hep-ph]}].

%%%%% NEW REFERENCES
\bibitem{su3}
G.~A.~Miller, H.~F.~Blichfeldt, and L.~E.~Dickson,
\textit{Theory and applications of finite groups},
John Wiley \& Sons, New York (1916);
\\
W.~M.~Fairbairn, T.~Fulton, and W.~H.~Klink,
\textit{Finite and disconnected subgroups of $SU_3$
and their application to the elementary-particle spectrum},
J.\ Math.\ Phys.\ {\bf 5} (1964) 1038.

\bibitem{6n2}
J.~A.~Escobar and C.~Luhn,
\textit{The flavor group $\Delta(6 n^2)$},
J.\ Math.\ Phys.\ {\bf 50} (2009) 013524
[{\tt hep-th/0809.0639}].

\bibitem{ludl1}
W.~Grimus and P.~O.~Ludl,
\textit{On the characterization of the $SU(3)$-subgroups of type C and D},
J.\ Phys.\ A {\bf 47} (2014) 075202
[{\tt arXiv:1310.3746 [math-ph]}].
%%%%%

\bibitem{varzielas}
I.~de Medeiros Varzielas and L.~Lavoura,
\textit{Golden ratio lepton mixing and nonzero reactor angle with $A_5$},
J.\ Phys.\ G {\bf 41} (2014) 055005
[{\tt arXiv:1312.0215 [hep-ph]}].

\bibitem{goldenratio}
Y.~Kajiyama, M.~Raidal, and A.~Strumia,
\textit{Golden ratio prediction for solar neutrino mixing},
Phys.\ Rev.\ D {\bf 76} (2007) 117301
[{\tt arXiv:0705.4559 [hep-ph]}];
\\
L.~L.~Everett and A.~J.~Stuart,
\textit{Icosahedral ($A_5$) family symmetry
and the golden ratio prediction for solar neutrino mixing},
Phys.\ Rev.\ D {\bf 79} (2009) 085005
[{\tt arXiv:0812.1057 [hep-ph]}].

\bibitem{tortola}
D.~V.~Forero, M.~T\'ortola, and J.~W.~F.~Valle,
\textit{Neutrino oscillations refitted},
Phys.\ Rev.\ D {\bf 90} (2014) 093006
[{\tt arXiv:1405.7540 [hep-ph]}].

\bibitem{fogli}
F.~Capozzi, G.~L.~Fogli, E.~Lisi, A.~Marrone, D.~Montanino, and A.~Palazzo,
\textit{Status of three-neutrino oscillation parameters, circa 2013},
Phys.\ Rev.\ D {\bf 89} (2014) 093018
[{\tt arXiv:1312.2878 [hep-ph]}].

\bibitem{schwetz}
M.~C.~Gonzalez-Garcia, M.~Maltoni, and T.~Schwetz,
\textit{Global analyses of neutrino oscillation experiments},
Nucl.\ Phys.\ B {\bf 908} (2016) 199
[{\tt arXiv:1512.06856 [hep-ph]}].

\bibitem{rodejohann}
W.~Rodejohann,
\textit{Unified parametrization for quark and lepton mixing angles},
Phys.\ Lett.\ B {\bf 671} (2009) 267
[{\tt arXiv:0810.5239 [hep-ph]}].

\bibitem{trimaximal}
See for instance W.~Grimus and L.~Lavoura,
\textit{A model for trimaximal lepton mixing},
J.\ High Energy Phys.\ {\bf 0809} (2008) 106
[{\tt arXiv:0809.0226 [hep-ph]}].

\bibitem{TM2}
C.~H.~Albright and W.~Rodejohann,
\textit{Comparing trimaximal mixing and its variants
with deviations from tri-bimaximal mixing},
Eur.\ Phys.\ J.\ C {\bf 62} (2009) 599
[{\tt arXiv:0812.0436 [hep-ph]}].

\bibitem{adulpravitchai}
See A.~Adulpravitchai, A.~Blum, and W.~Rodejohann,
\textit{Golden ratio prediction for solar neutrino mixing},
New J.\ Phys.\  {\bf 11} (2009) 063026
[{\tt arXiv:0903.0531 [hep-ph]}],
and the references therein.




\end{thebibliography}
\end{document}